\documentclass{article}
\usepackage{arxiv}
\usepackage[english]{babel}
\usepackage[utf8]{inputenc} 
\usepackage[T1]{fontenc}    
\usepackage{hyperref}       
\usepackage{url}            
\usepackage{booktabs}       
\usepackage{amsmath}
\usepackage{amsfonts}       
\usepackage{nicefrac}       
\usepackage{microtype}      
\usepackage{graphicx}
\usepackage[square,numbers]{natbib}
\usepackage{doi}
\usepackage{tabularx}
\usepackage{multirow}
\usepackage{float}
\usepackage[justification=centering]{caption}
\usepackage{subcaption}
\usepackage{flafter}
\usepackage{threeparttable}

\title{Machine learning-based detection of cardiovascular disease using ECG signals: performance vs. complexity}

\author{Huy PHAM \\
	HSE University \\ 
	\texttt{nfam\_2@edu.hse.ru} \\
	\And
	Konstantin EGOROV \\
	Sber AI Lab\\
	\texttt{egorov.k.ser@sberbank.ru} \\
        \And
        Alexey KAZAKOV \\
        Sber AI Lab \\
        \texttt{AAleksanKazakov@sberbank.ru} \\
        \And
        Semen BUDENNYY* \\
        Sber AI Lab \\
        Artificial Intelligence Research Institute (AIRI) \\
        \texttt{sanbudenny@sberbank.ru} \\
        *Corresponding author\\
}

\renewcommand{\shorttitle}{Machine learning-based detection of cardiovascular disease using ECG signals: performance vs. complexity}

\hypersetup{
pdftitle={Machine learning-based detection of cardiovascular disease using ECG signals: performance vs. complexity},
pdfsubject={cs.LG},
pdfauthor={Huy Ngoc Pham, Konstantin Egorov, Alexey Kazakov, Semen Budennyy},
pdfkeywords={ECG signal, Cardiovascular disease, Arrhythmia, Deep Learning, CinC, Poincaré diagram},
}

\begin{document}
\maketitle

\begin{abstract}
	Cardiovascular disease remains a significant problem in modern society. Among non-invasive techniques, the electrocardiogram (ECG) is one of the most reliable methods for detecting abnormalities in cardiac activities. However, ECG interpretation requires expert knowledge and it is time-consuming. Developing a novel method to detect the disease early could prevent death and complication. The paper presents novel various approaches for classifying cardiac diseases from ECG recordings. The first approach suggests the Poincaré representation of ECG signal and deep-learning-based image classifiers (ResNet50 and DenseNet121 were learned over Poincaré diagrams), which showed decent performance in predicting AF (atrial fibrillation) but not other types of arrhythmia. XGBoost, a gradient-boosting model, showed an acceptable performance in long-term data but had a long inference time due to highly-consuming calculation wihtin the pre-processing phase. Finally, the 1D convolutional model, specifically the 1D ResNet, showed the best results in both studied CinC 2017 and CinC 2020 datasets, reaching the F1 score of 85\% and 71\%, respectively, and that was superior to the first-ranking solution of each challenge. The paper also investigated efficiency metrics such as power consumption and equivalent CO2 emissions, with one-dimensional models like 1D CNN and 1D ResNet being the most energy efficient. Model interpretation analysis showed that the DenseNet detected AF using heart rate variability while the 1DResNet assessed AF pattern in raw ECG signals. 

\end{abstract}


\section{Introduction}
Cardiovascular disease is a serious public health problem that affects millions of people worldwide and is also a leading cause of death~\cite{Tsao2022}. The expense of healthcare, lost productivity, and a diminished quality of life due to heart illness has a significant economic and social impact on individuals, families, and society as a whole~\cite{mensah2019global}. This emphasizes the value of early disease identification. While the electrocardiogram (ECG) is considered as the most crucial method for detecting and diagnosing cardiac problem~\cite{Kligfield2007}, it takes time and requires trained professionals with specialized skills to interpret ECGs. The ECG analysis task includes beat annotation and signal classification. While the former deals with aligning the signal segment to the heart contraction, the latter tries to predict the disease from the signal data.

In the domain of ECG classification, there are a number of methods ranging from feature-based models to deep-learning based ones. The feature-based approach takes advantage of a feature extraction technique and a machine learning model. The methods for feature extraction are very diverse, however, the domain-dependent features, statistical descriptors, morphological characteristics, and frequency-based features are widely chosen \cite{Selcan2018}.

Challenges (with annotated datasets provided) of ECG classification such as The PhysioNet/Computing in Cardiology Challenge (CinC) 2017 and 2020~\cite{cinc2017, cinc2020} are aimed to provide opportunities for data science community to develop a novel method for automatic detection. While CinC 2017 focused on arrhythmia disease, CinC 2020 contained ECG signals in a wide range of cardiac abnormalities. Although there are many efforts to apply machine learning/deep learning approaches to reach the highest performance, the results were still modest, especially in CinC 2020.

Additionally, there are state-of-the-art methods that improve the performance of signal classification, especially in deep-learning-based methods. Besides the improvement of accuracy, the architecture of models becomes more complicated, so they require more energy to train and have a long inference time. This problem limits the application of the method, especially in handheld and wearable devices.

In this study, we focused on enhancing ESG classification approaches in terms of performance, numerical complexity, inference time, and its interpretability.

\noindent \textbf{Contribution}. The contribution of our paper is threefold: 

\begin{itemize}
\item First, we introduce a pipeline for heart disease classifier evaluation in terms of performance and numerical complexity. 
\item Second, we achieved a state-of-the-art level (regards to CinC 2017 and CinC 2020 challenges) performance with 1D ResNet model for both CinC 2017 and CinC 2020 benchmarks.
\item Third, we provided interpretation techniques for DenseNet121 and 1D ResNet models.

\end{itemize}

\section{Related work}
Many previous works on ECG classification focus on detecting heart disease, including atrial fibrillation, tachycardia, bradycardia, arrhythmia, and other problems \cite{Hong2020, Ribeiro2020, Zhang2021}. Some other studies try to predict mortality rates or demographic characteristics \cite{Attia2019, Lima2021}. The input of these models is usually raw ECG signal in single or multiple leads; however, sometimes only ECG images were used \cite{Jun2018}. 

The data could be transformed into good features before feeding to a machine learning model or be processed automatically to produce a high dimensional representation by a deep learning model. In the feature-based method, four groups of descriptors can be extracted from the ECG signal: time-domain features, nonlinear-domain features, distance-based features, and time-series features (\cite{Bertsimas2021}). In the next step, the classifier, such as logistics regression, support vector machine, or boosting algorithms, gives the prediction based on these features. Besides that, the deep learning-based approaches perform feature extracting and predicting simultaneously. These models take advantage of multiple layer perceptron \cite{Ribeiro2020}, CNN model \cite{Hong2020}, or LSTM model \cite{Corradi2019} for extracting the high-level features of ECG signal.


The work by Jun et al. \cite{Jun2018} focused on predicting arrhythmia diseases based on ECG beat. The author used the data from the MIT-BIH arrhythmia database. In the pre-processing phase, the ECG signal was partitioned and centered based on the Q-wave peak time before being plotted to generate a 128 x 128 grayscale image as the input for learning. Data augmentation was performed by cropping and resizing the training data images. The CNN model tried to classify eight labels from these plotting. In that paper, the AlexNet, VGGNet, and a customized CNN architecture were used to optimize the performance. The proposed model reached 0.989 AUC and over 99\% accuracy. The data augmentation showed the benefit of raising the sensitivity of the models. Compared to our approach on the Poincaré diagram, we used the scatter plot on heart rate variation instead of the raw signal. The used CNN architectures were the novel models, including ResNet and DenseNet. The cropping and resizing augmentation were not applied because this technique could modify the dispersion of the Poincaré plot, which leads to the wrong prediction. Instead, the limited random erasing on the input image was used to regularize the CNN model.

There was a concerted effort by Shenda Hong et al. \cite{Hong2020} to develop an ensemble system to process the waveform data. Hong’s architecture includes three key parts named the model zoo, the ensemble composer, and the real-time serving system. The model zoo takes the responsibility for training the data with several collections of hyperparameters, while the ensemble composer would figure out the best set of models under the constraints of validation performance and latency. The serving system takes the output of the ensemble composer and then deploys a system that can handle massive input data as well as queries in real time. The authors performed many experiments on signal leads with several types of deep learning models such as CNN, ResNet, ResNeXt, and RegNet. The experiments showed that this system could reach an accuracy of 95\% and a latency of under a second on a 64-bed simulation.

In the study by Ribeiro et al. \cite{Ribeiro2020}, the authors investigate how to train a one-dimensional convolutional neural network to predict cardiac diseases. The trained data includes more than two million 12-lead ECG signals that are between 7 to 10 seconds. The dataset was annotated semi-automatically that combines algorithms and human verification. The chosen model was based on the ordinary 1D ResNet architecture that has 4 Residual Blocks with kernel size increasing from 128 to 320. Finally, this pipeline surpassed human performance with an F1 score of over 80\%.

The work by Zhang’s team \cite{Zhang2021} proved the dominant results of deep learning compared to the feature-based machine learning model. In this study, 1D CNN was trained on 12-lead ECG recordings from CPSC 2018 database. The model could predict 9 subtypes of arrhythmias with an F1 score of over 80\%. The impressive idea of this work is to use the SHAP value to explain the model output at the individual level as well as the population level. At the individual level, the model could show the characteristics of ECG that support the model decision such as the abnormal QRS pattern following the P waves in AF, or the prolongated PR distance in IAVB. At the population level, the authors examined the contribution of each lead to model output, so that the lead II, aVR, V1, V2, V5, and V6 are the most important leads in their model.

Besides the signal classification problem, the beat-level annotation is also investigated. Corradi et al. \cite{Corradi2019} took advantage of recurrent neural networks to annotate the ECG with over 90\% accuracy in many public datasets. Their model was also efficient enough to deploy in the wearable device. In 2020, Teplitzky presented BeatLogic.~\cite{Teplitzky2020} This was a comprehensive system that could detect and classify the cardiac beat and rhythm simultaneously by using the 1D-CNN-based model. As a result, BeatLogic outperformed other methods on every mentioned task.  

Our proposed method also takes advantage of the Poincaré plot, a standard procedure for studying heart rate variability. Early work by \cite{Park2009} used point coordinates in the Poincaré plot to calculate the interval and variability of interbeat before using a support vector machine model to classify AF and non-AF patients. \cite{Lian2011} combined the RR interval and the difference of RR intervals to make a robust AF classifier with few heartbeats. \cite{Zhang2015} deploys the ensemble of neural networks to extract five geometric patterns in the Poincaré plot, including comet, torpedo, fan, double side lobe, and multiple side lobes, before using them to classify the major cardiac arrhythmias. \cite{Bashar2021} proposed a modified Poincaré plot from heart rate difference. The features extracted from these plots by image processing help diagnose AF from Premature Atrial/Ventricular Contraction.


\section{Methods}
\subsection{Data}
The data was collected from two challenges PhysioNet/CinC Challenge 2017 and 2020~\cite{cinc2017, cinc2020}. The disclosure data was split into train/validation/test subsets with the ratio 60/20/20.

\begin{table}[h]
\caption{Descriptive statistics for CinC 2017 and CinC 2020 datasets}
\centering
\begin{tabular}{llrrrrrr}
\hline
\multirow{2}{*}{\textbf{Dataset}} & \multirow{2}{*}{\textbf{Dataset}} & \multicolumn{1}{l}{\multirow{2}{*}{\textbf{Samples}}} & \multicolumn{1}{l}{\multirow{2}{*}{\textbf{Rate}}} & \multicolumn{4}{c}{\textbf{Length (second)}} \\ \cline{5-8} 
 &  & \multicolumn{1}{l}{} & \multicolumn{1}{l}{} & \multicolumn{1}{l}{\textbf{Mean}} & \multicolumn{1}{l}{\textbf{Min}} & \multicolumn{1}{l}{\textbf{Median}} & \multicolumn{1}{l}{\textbf{Max}} \\ \hline
 & Train & 5116 & 300 & 32.4 & 9.1 & 30.0 & 61.0 \\
CinC 2017 & Val & 1706 & 300 & 32.8 & 9.8 & 30.0 & 60.6 \\
 & Test & 1706 & 300 & 32.5 & 9.0 & 30.0 & 60.8 \\ \hline
 & Train & 25860 & 257 - 1000 & 15.4 & 5.0 & 10.0 & 1800.0 \\
CinC 2020 & Val & 8620 & 257 - 1000 & 16.1 & 5.0 & 10.0 & 1800.0 \\
 & Test & 8621 & 257 - 1000 & 16.1 & 5.0 & 10.0 & 1800.0 \\ \hline
\end{tabular}
\end{table}

The CinC 2017 dataset was recorded by AliveCor device and contains 8528 single lead signals. The length of recordings is from 9 to 60 seconds, the average length is about 32 seconds. Every ECG signal was recorded at 300 Hz and already filtered by the recorder. The host provided the data in WFDB format with a .mat file containing signal data and a .hea file containing headers for basic information including ID, recording parameters, and patient information. 

\begin{figure}[h]
    \centering
    \includegraphics[width=\textwidth]{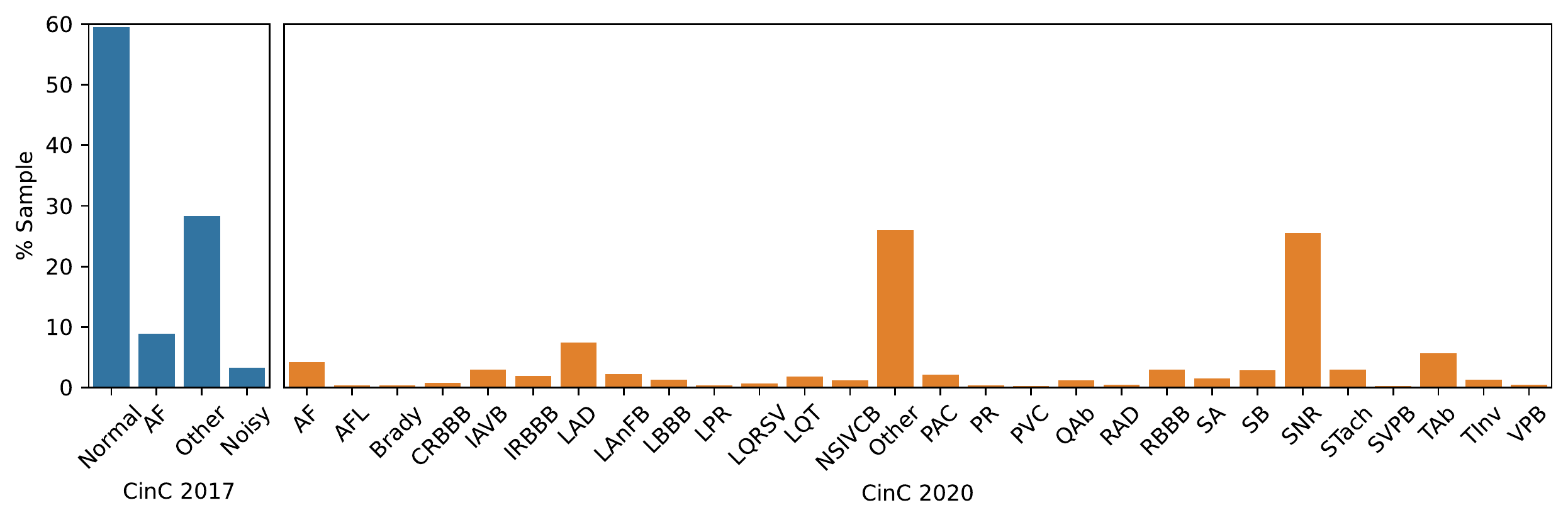}
    \caption{The classes distribution in datasets CinC 2017 (left side) and CinC 2020 (right side)}
    \label{fig:classes_distribution}
\end{figure}

The CinC 2020 dataset contains 12-lead signals which come from five different sources: CPSC Database and CPSC-Extra Database, INCART Database, PTB and PTB-XL Database, The Georgia 12-lead ECG Challenge (G12EC) Database, and the Private Database.

The dataset CPSC is the collection of the ECG signals of Chinese patients which were recorded at 500 Hz. The patient's gender and age were disclosed in this dataset, however, the age of over-89-year-old patients is masked as 92 due to the HIPAA guidelines.
The INCART database contains 30-min recordings at 257 Hz while the PTB and Georgia datasets consist of 10-second recordings only.
The private data is not public so this source was not     included in our work.
The remaining dataset was split into train/test/split with the ratio 60/20/20.
Like the CinC 2017 dataset, the data from CinC 2020 is also WFDB-compliant. The header files embedded the demographics information and diagnosis labels. 

\subsection{Model architecture and training pipeline}

\begin{figure}[h]
    \centering
    \includegraphics[width=\textwidth]{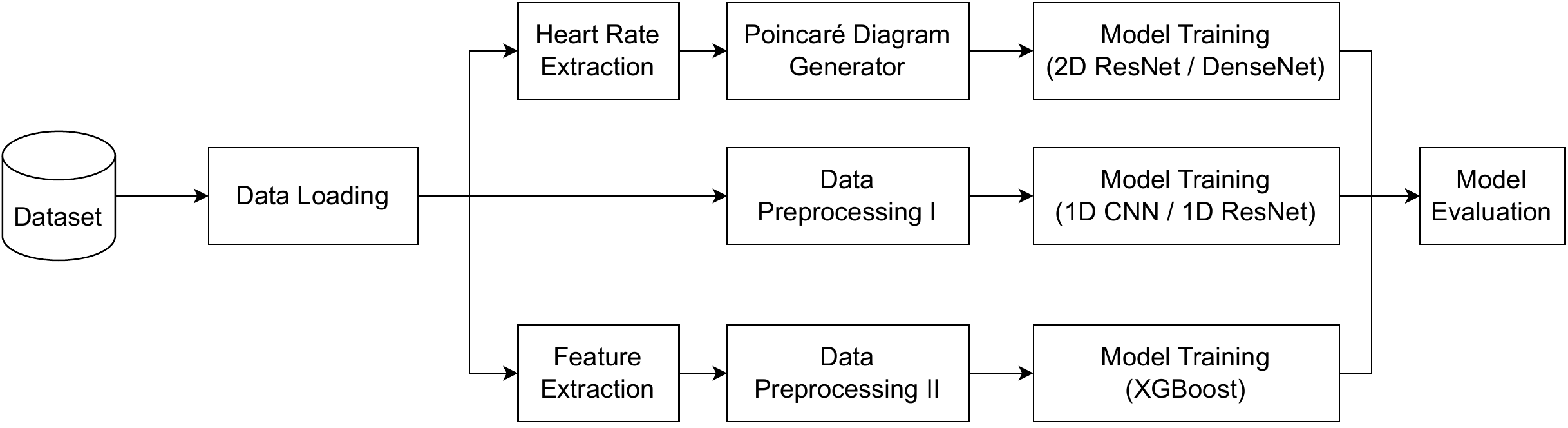}
    \caption{The high-level scheme of training pipeline.}
    \label{fig:pipeline}
\end{figure}

This section describes the detail of the configuration of each model as well as the flow of data when training the model. The overview of the training pipeline is given in Figure \ref{fig:pipeline} and interpreted in the following.

\subsubsection{Learning over Poincaré representation}

\begin{figure}[h]
    \centering
    \begin{subfigure}[t]{0.45\textwidth}
        \centering
        \includegraphics[width=\textwidth]{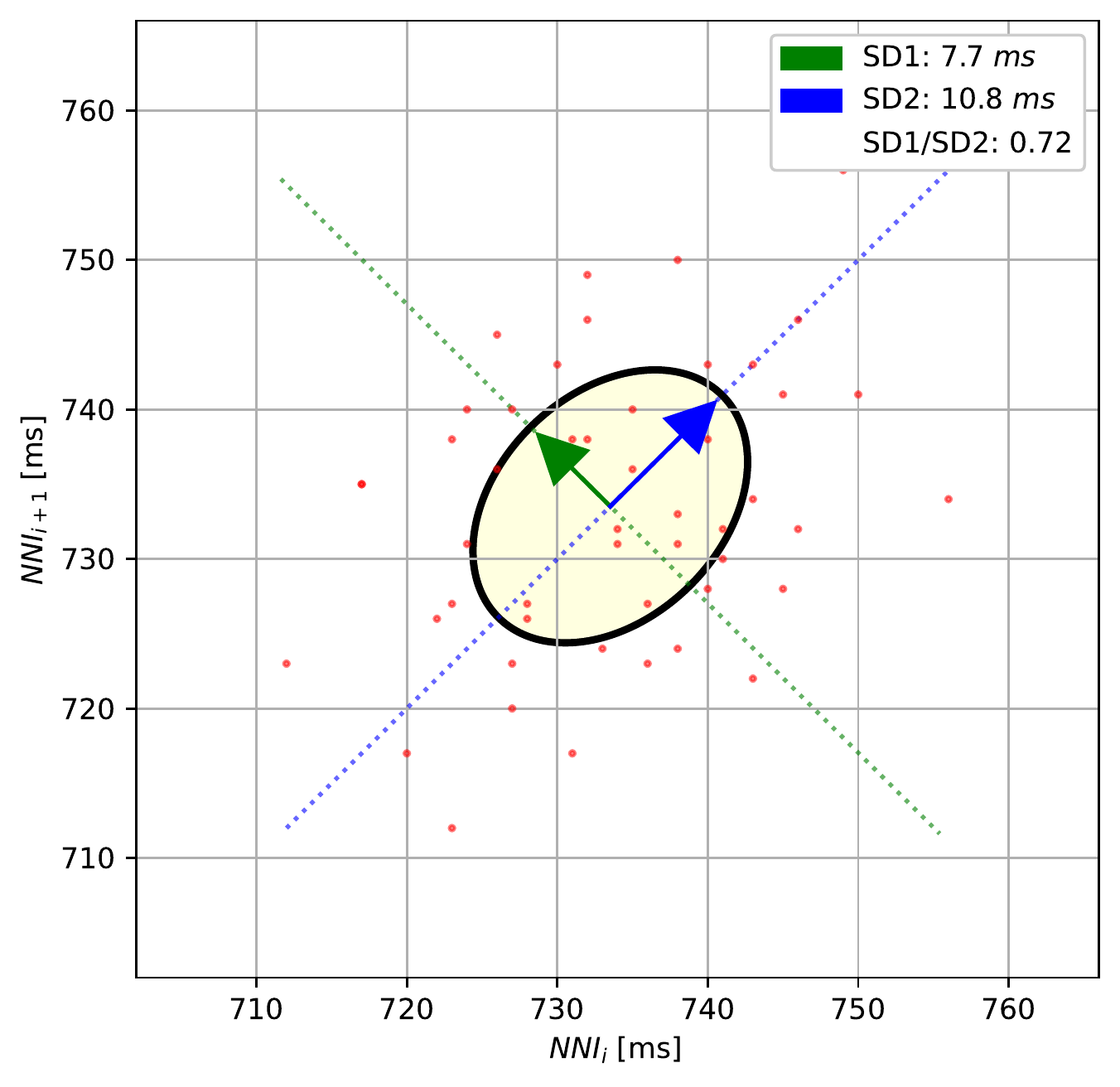}
        \caption{}
        \label{fig:short_poincare}
    \end{subfigure}%
    \hfill
    \begin{subfigure}[t]{0.45\textwidth}
        \centering
        \includegraphics[width=\textwidth]{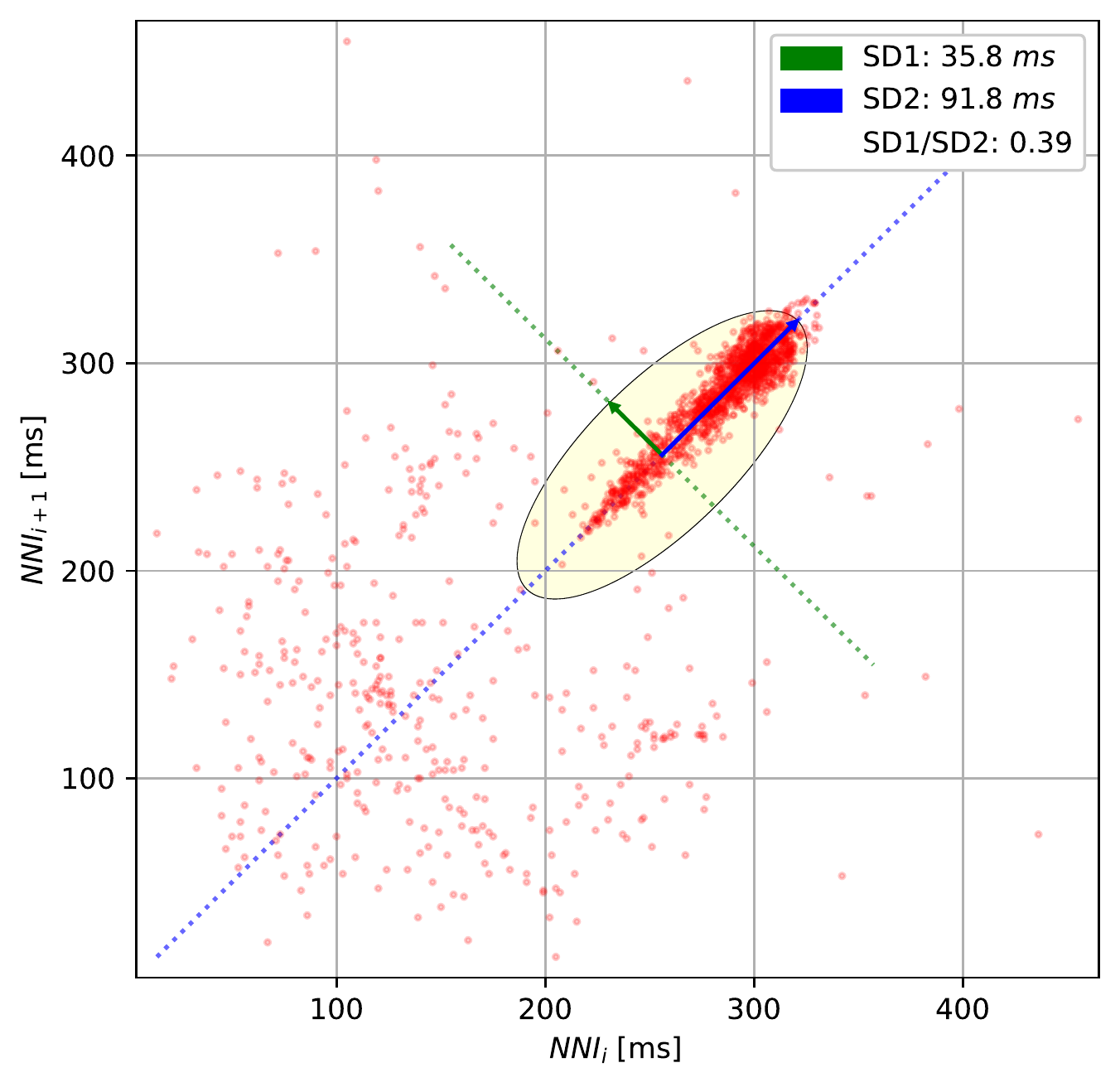}
        \caption{}
        \label{fig:long_poincare}
    \end{subfigure}
    \caption{The Poincaré diagrams of the short-term (a) and long-term (b) ECG. The diagrams plot the normal R-peak intervals (or NN~intervals).}
    \label{fig:poincare_diagram}
\end{figure}

For the methods based on the Poincaré diagram, the input ECG signals were preprocessed by \texttt{biosppy}~\cite{biosppypaper} to extract the R-peak positions from the signal \cite{zong2003open}. This library filters the ECG signal in the frequency range from 3 to 45 before using Hamilton algorithm \cite{hamilton2002open} to detect the R-peak. The distance between R-peaks (or RR intervals) was evaluated from the R-peak location. Furthermore, in our study, we only used the NN intervals which are the distances between normal R-peaks after removing the noise and artifacts. The Poincaré diagram was constructed by plotting the scatter charts for $NN_i$ and $NN_{i+1}$ intervals. Figure \ref{fig:poincare_diagram} shows the examples of Poincaré diagram of a short and long recording.

To predict the heart disease over the Poincaré diagram, the default architecture of \texttt{ResNet50}~\cite{he2016deep} and \texttt{DenseNet121}~\cite{huang2017densely} were used to train from the scratch (without pre-trained weights). The last layers of these models were also tailored to match the number of classes of each dataset. The Gradient-weighted Class Activation Mapping (GradCAM) \cite{Selvaraju2017} was constructed to explore the mechanism behind the model decision.

\subsubsection{Learning over 1D signal}

The 1D CNN model comprises twelve base blocks. Each base block consists of a 1D Convolutional layer, 1D Batch Normalization, Activation function, Pooling layer, and Drop-out layer.  In the 1D Convolutional layer, the padding is always \texttt{'valid'} while the stride size is always 1. The output channel starts at 256 and decreases gradually to 32 in the last convolutional layer, and the kernel size starts at 20 followed by 5 layers with a kernel size of 5, and then 3 for the remaining layers. The Batch Normalization layers have the number of weights the same as the number of output channels of the prior convolutional layer. The momentum of normalization is 0.99 for every block. The Pooling of base block is \texttt{MaxPooling1d} of which the kernel size and stride size are 2. The dropout probability is set to 0.3 in every place. Before flattening the tensor and feeding to the last fully connected layer for the logit outputs, there is an Average pooling layer with a kernel size of 1 and stride size of 2.

The structure of the base block in 1D ResNet includes a 1D Convolutional layer, 1D Batch Normalization, ReLU activation function, a Drop-out layer, another 1D Convolutional layer, and 1D Batch Normalization. In a base block, the input would go through these layers before adding the residual which is also the input tensor. This summation is activated by the ReLu function after leaving the block.

In our work, the 1D ResNet starts with a 1D Convolutional layer with a kernel size of 15 and the number of output channels is 64 followed by a 1D Batch Normalization, ReLU Activation function, and Max Pooling layer. After that, there are four base blocks with kernel sizes increasing from 65 to 256. The output of the last base block goes through two pooling layers: an Average Pooling layer and a Max Pooling layer. These outputs are concatenated before feeding to the final fully connected layer to compute the output logits.

In both 1D CNN and 1D ResNet models, the signals are converted to first-order difference and scaled to zero mean and unit variance before transferring to the models. We also adapted the GradCAM \cite{Selvaraju2017} to figure out which regions in ECG recordings contribute to the model results.

\subsubsection{Learning over XGBoost feature space}

In the pipeline of the XGBoost model, the processed ECG signals need to feed to module \texttt{tsfresh} to extract the features before training model. The features extraction used the default setting in the subset \texttt{EfficientFCParameters} but filtered out the time-consuming features including: \texttt{approximate\_entropy}, \texttt{sample\_entropy}, \texttt{matrix\_profile}, \texttt{number\_cwt\_peaks}, \texttt{partial\_autocorrelation}, \texttt{agg\_linear\_trend}, \texttt{augmented\_dickey\_fuller}. In the feature matrix, the pipeline filled the missing data with $-999$ and removed the low-variance features.

The hyperparameters of \texttt{XGBoost} were optimized by searching within the predefined space (Figure \ref{tab:xgb_hp}). The optimum collection was found by Bayesian optimization implemented in the library \texttt{scikit-optimize}.~\cite{skopt2021} The number of search trials was limited to 100 because of time constraints.

\begin{table}[!ht]
   \caption{The hyperparameters searching space of XGBoost.} 
   \label{tab:xgb_hp}
   \small
   \centering
   \begin{tabular}{p{0.15\linewidth} | p{0.22\linewidth}  p{0.15\linewidth}  p{0.11\linewidth}}
   \toprule\toprule
   \textbf{Component} & \textbf{Hyperparameter} & \textbf{Range} & \textbf{Distribution} \\ 
   \midrule

   Feature Elimination & \texttt{min\_features\_to\_select} & $[10, \text{\# features}]$ & uniform \\ 
   
    XGBClassifier & \texttt{max\_depth} & $[2, 100]$ & uniform \\ 
    XGBClassifier & \texttt{gamma} & $[10^{-3} , 10^3]$ & log-uniform \\ 
    XGBClassifier & \texttt{eta} & $[10^{-3} , 10^3]$ & log-uniform \\ 
    XGBClassifier & \texttt{scale\_pos\_weight} & $[10^{-3} , 10^3]$ & log-uniform \\ 
    XGBClassifier & \texttt{reg\_lambda} & $[10^{-3} , 10^3]$ & log-uniform \\ 
    XGBClassifier & \texttt{reg\_alpha} & $[10^{-3} , 10^3]$ & log-uniform \\ 
   
   \bottomrule
  \end{tabular}
  
\end{table}

\section{Results}
\subsection{Cardiovascular diseases classification}

\begin{figure}[h]
    \centering
    \includegraphics[width=0.95\textwidth]{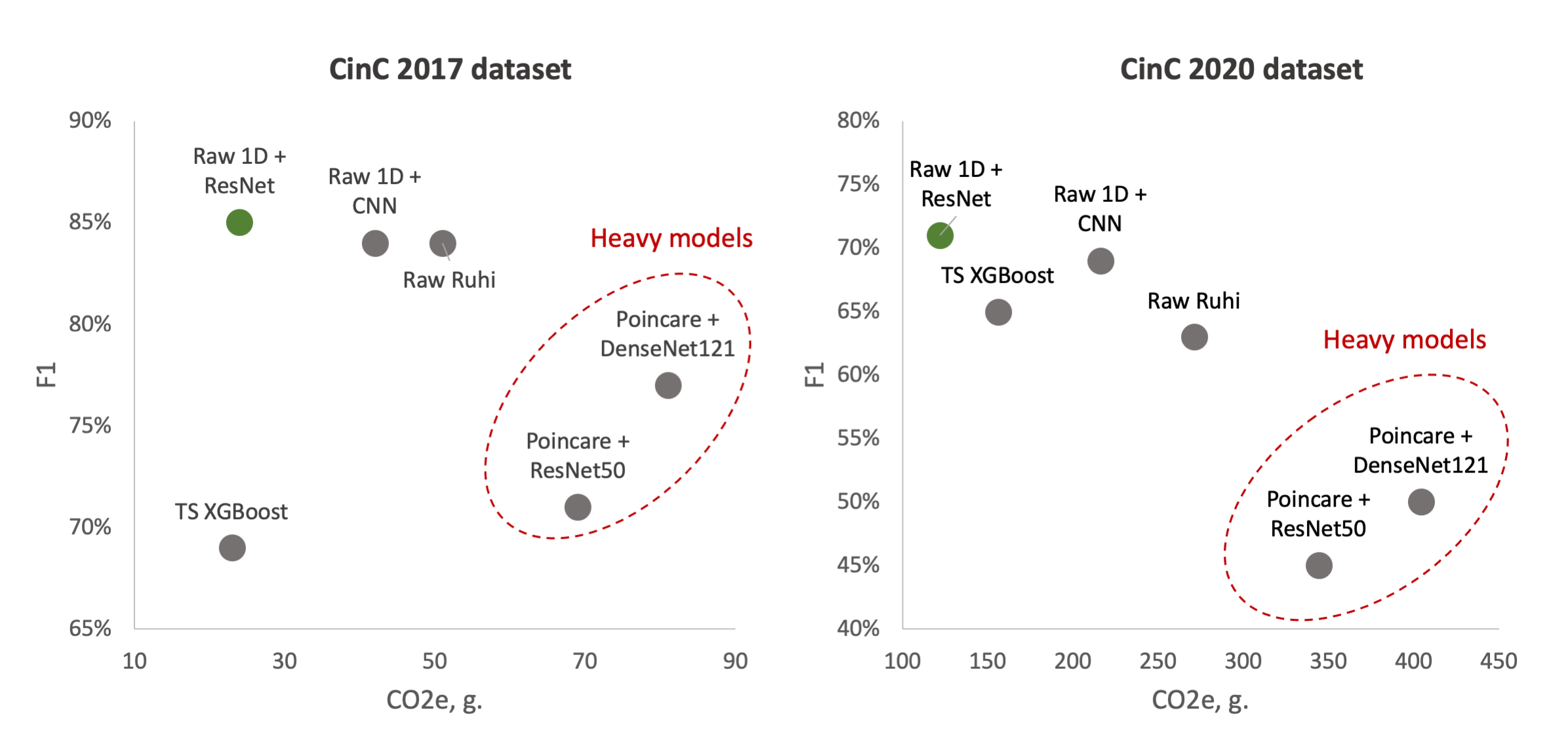}
    \caption{F1 score vs. $CO_2$ emissions: left side - models learned over CinC 2017 dataset, right side - models learned over CinC 2020 dataset. Dotted red ellipses highlight relatively heavy models}
    \label{fig:co2_vs_f1}
\end{figure}

The experiment results showed the superior performance of the 1D ResNet model learned over raw data in both datasets. Especially, in CinC~2020, this model surpassed the 1st rank solution by a large margin. The comparison of F1 scores and the efficiency metrics (power consumption, eq. CO2) are given in Table \ref{tab:scores}. To better illustrate the tandem "Perfomance vs. Complexity" for examined models the figure \ref{fig:co2_vs_f1} fives cross-plots on F1 score and CO2 emissions for both datasets. In particular, one can reveal that DenseNet121 and ResNet50 models learned over Poincare diagrams stand out from other models as inefficient while ResNet learned on raw ECG signals outperforms. 

\begin{table}[!ht]
    \caption{Performance on test datasets for CinC~2017 and CinC~2020 competitions} 
    \label{tab:scores}
    \small
    \centering
    \begin{threeparttable}
    \begin{tabular}{p{0.1\linewidth} | p{0.1\linewidth}  p{0.11\linewidth} p{0.1\linewidth} p{0.17\linewidth} p{0.12\linewidth} }
    \toprule\toprule
    \textbf{Dataset} & \textbf{Input data} & \textbf{Model} & \textbf{F1 score} 
    & \textbf{Power consumption, Wh} 
    & \textbf{$CO_2$, g}  \\ 
    \midrule
    
            & Poincaré     & ResNet50       & 0.71          & 127 & 69         \\ 
                     & Poincaré     & DenseNet121    & 0.77          & 148 & 81         \\ 
    CinC~2017       & Raw Signal   & 1D CNN         & 0.84          & 077 & 42   \\ 
            & Raw Signal   & 1D ResNet      & \textbf{0.85} & 44 & 24 \\ 
           & Raw Signal   & Ruhi*           & 0.84          & 92 & 51 \\ 
           & Time Series  & XGBoost        & 0.69          & \textbf{42} & \textbf{23} \\ \hline
    
          & Poincaré     & ResNet50       & 0.45          & 630 & 344 \\ 
            & Poincaré     & DenseNet121    & 0.50          & 740 & 404 \\ 
    CinC~2020          & Raw Signal   & 1D CNN         & 0.69          & 396 & 216 \\ 
           & Raw Signal   & 1D ResNet      & \textbf{0.71} & \textbf{223} & \textbf{122} \\ 
           & Raw Signal   & PRNA*          & 0.63          & 497 & 271 \\ 
           & Time Series  & XGBoost        & 0.65          & 286 & 156 \\
    \bottomrule
    \end{tabular}
    \begin{tablenotes}
     \item *The first rank solution.
    \end{tablenotes}
    \end{threeparttable}
\end{table}

The Poincaré-based methods have adequate performance in the CinC~2017 challenge. However, they do not perform well in the CinC~2020 challenge. In particular, some classes are not discriminated in the Poincaré diagrams. The models ResNet50 and DenseNet121 only identified the types AF, SB, SNR, STach and other, while the metrics for the remaining types are close to zero. This result is understandable as the information on heart rate variability is not sufficient to identify many types of heart disease. They are also the most power-hungry models: ResNet50 and DenseNet121 consumed 2 to 3 times more energy than the others.

The XGBoost is under-expected because it ranked lowest in CinC~2017 and only third in CinC~2020 despite the gradient-boosting family usually gaining the highest place at many machine learning benchmarks. In terms of power consumption, this model is very efficient when processing the short-term signal in CinC~2017; however, the required energy increases by seven folds when processing the long-term signals in CinC~2020. This phenomenon is the result of the heavy preprocessing step in this pipeline.

The unidimensional convolutional models yielded excellent results in both classification performance and energy usage. The 1D CNN and 1D ResNet shared the top~2 places in both datasets. In CinC 2020, the 1D ResNet is the best model, followed by the 1D CNN with only 1\% lower. The former is also very energy efficient: the 1D ResNet consumed more energy than only XGBoost in CinC 2017 and was the most efficient model in CinC 2020. Despite the 1D CNN having a simpler base block than the 1D ResNet, the CNN required more layers than ResNet (12 CNN blocks versus 4 ResNet blocks). So the former required more power and performed less well than the latter.


\begin{table}[h]
    \caption{The F1 score on each sources in CinC~2020 dataset.}
    \label{tab:cinc2020_f1_by_source}
    \centering
    \begin{threeparttable}
    \begin{tabular}{l|cccccc}
    \toprule \toprule
    \textbf{Model} 
        & \begin{tabular}{@{}c@{}}\textbf{G12EC} \\ $\bar{l}=10$ \end{tabular}
        & \begin{tabular}{@{}c@{}}\textbf{PTB-XL} \\ $\bar{l}=10$ \end{tabular}
        & \begin{tabular}{@{}c@{}}\textbf{CPSC} \\ $\bar{l}=16$ \end{tabular} 
        & \begin{tabular}{@{}c@{}}\textbf{CPSC-Extra} \\ $\bar{l}=16$ \end{tabular}
        & \begin{tabular}{@{}c@{}}\textbf{PTB} \\ $\bar{l}=109$ \end{tabular} 
        & \begin{tabular}{@{}c@{}}\textbf{INCART} \\ $\bar{l}=1800$ \end{tabular} \\ \midrule
    ResNet50 & 0.28 & 0.56 & 0.35 & 0.33 & 0.16 & 0.20 \\ 
    DenseNet121 & 0.36 & 0.58 & 0.38 & 0.50 & 0.77 & 0.60 \\ 
    1D CNN & 0.55 & \textbf{0.76} & 0.61 & 0.68 & 0.85 & \textbf{0.74} \\ 
    1D ResNet & \textbf{0.59} & \textbf{0.76} & \textbf{0.70} & 0.66 & 0.85 & 0.70 \\
    XGBoost & 0.56 & 0.69 & 0.55 & \textbf{0.77} &\textbf{ 0.87} & \textbf{0.74} \\
    \bottomrule
    \end{tabular}
    \begin{tablenotes}
        \item $\bar{l}$ is the average length of signal in seconds
    \end{tablenotes}
    \end{threeparttable}
\end{table}

We also analyzed the performance of investigated models in each source of the CinC 2020 dataset (Table \ref{tab:cinc2020_f1_by_source}). The ResNet50 was good at the short-term recordings while performing poorly in long-term data. The DenseNet121 was better than ResNet50 in long-term signal classification but did not surpass the 1D Convolutional model. The XGBoost outperformed the others in long-term ECG. However, the number of long-term signals is modest, so their metrics might not stable.

\subsection{Models interpretation}

\subsubsection{DenseNet121 on Poincaré diagram classification}
\begin{figure}[h]
    \centering
    \includegraphics[width=0.85\textwidth]{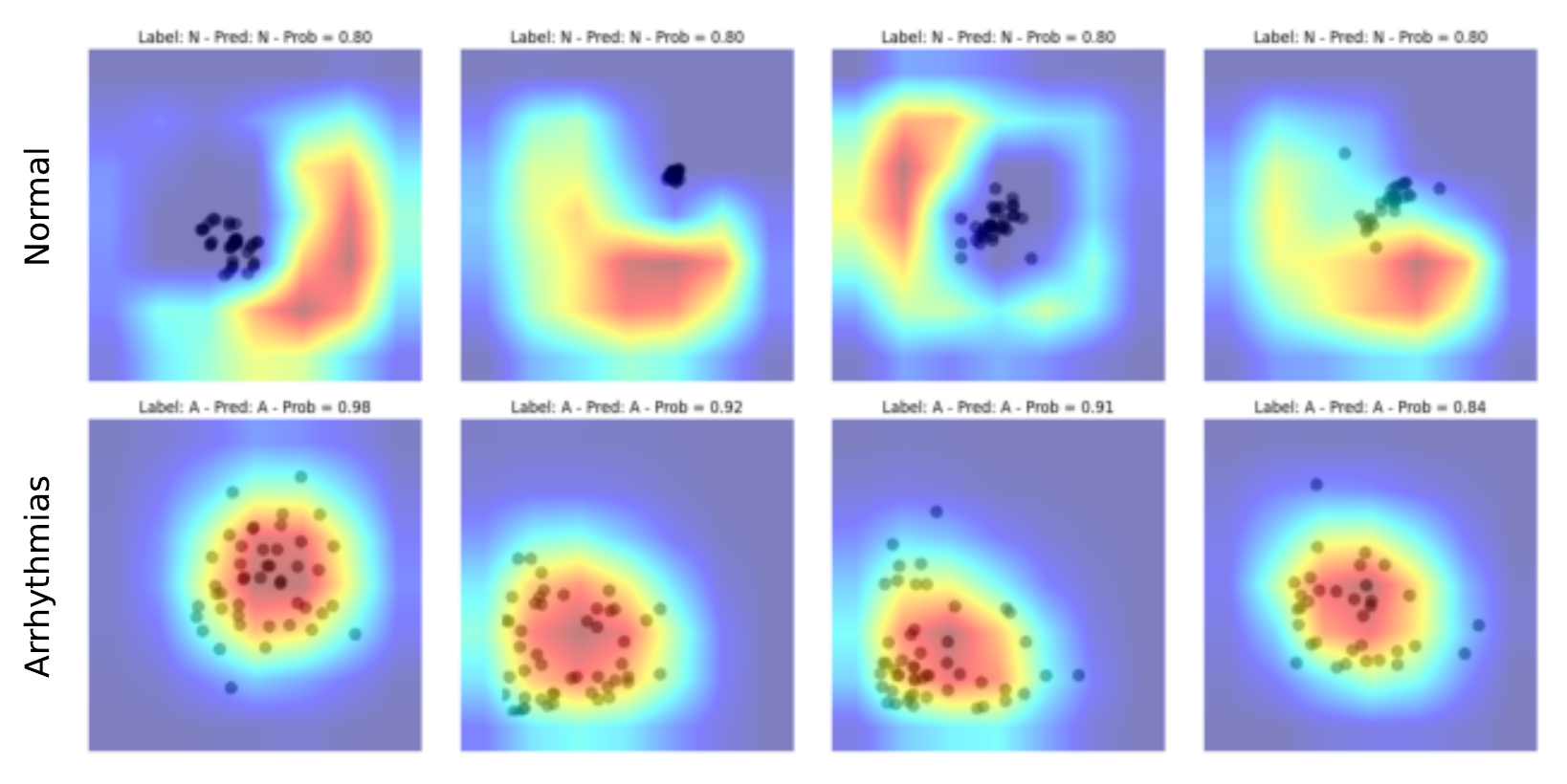}
    \caption{Explaining CinC~2017 predictions on the Poincaré diagrams using GradCam}
    \label{fig:explain_densenet}
\end{figure}

Figure \ref{fig:explain_densenet} visualized the GradCAM output of DenseNet121 on CinC 2017. We can see how this model processes the Poincaré diagram differently. In the Normal graph, the model focused on the area in the upper-left and lower-right, while the shape of the point cloud was ignored. In the arrhythmia diagram, the model focuses on the point cloud or the diversion of data. 

This behavior of the model is compatible with human knowledge. For ordinary people, we do not expect any data point far away from the diagonal of the diagram. Any data point in the upper left or lower right area is evidence of abnormal changes in heart rate and predicts the problem.
While in arrhythmia patients, because of the fluctuation in the heartbeat statistics, the data should be very varied and form a spreading cloud in the Poincaré diagram. The bigger cloud shows more variation in heart rate.

\subsubsection{1D ResNet on ECG signal classification}

Our work also took advantage of GradCAM to explore the mechanism of the 1D ResNet model. In medical literature, the ECG of Atrial fibrillation was detected by the irregular pattern in P- and T-waves around the QRS complex.

\begin{figure}[h]
    \centering
    \includegraphics[width=0.65\textwidth]{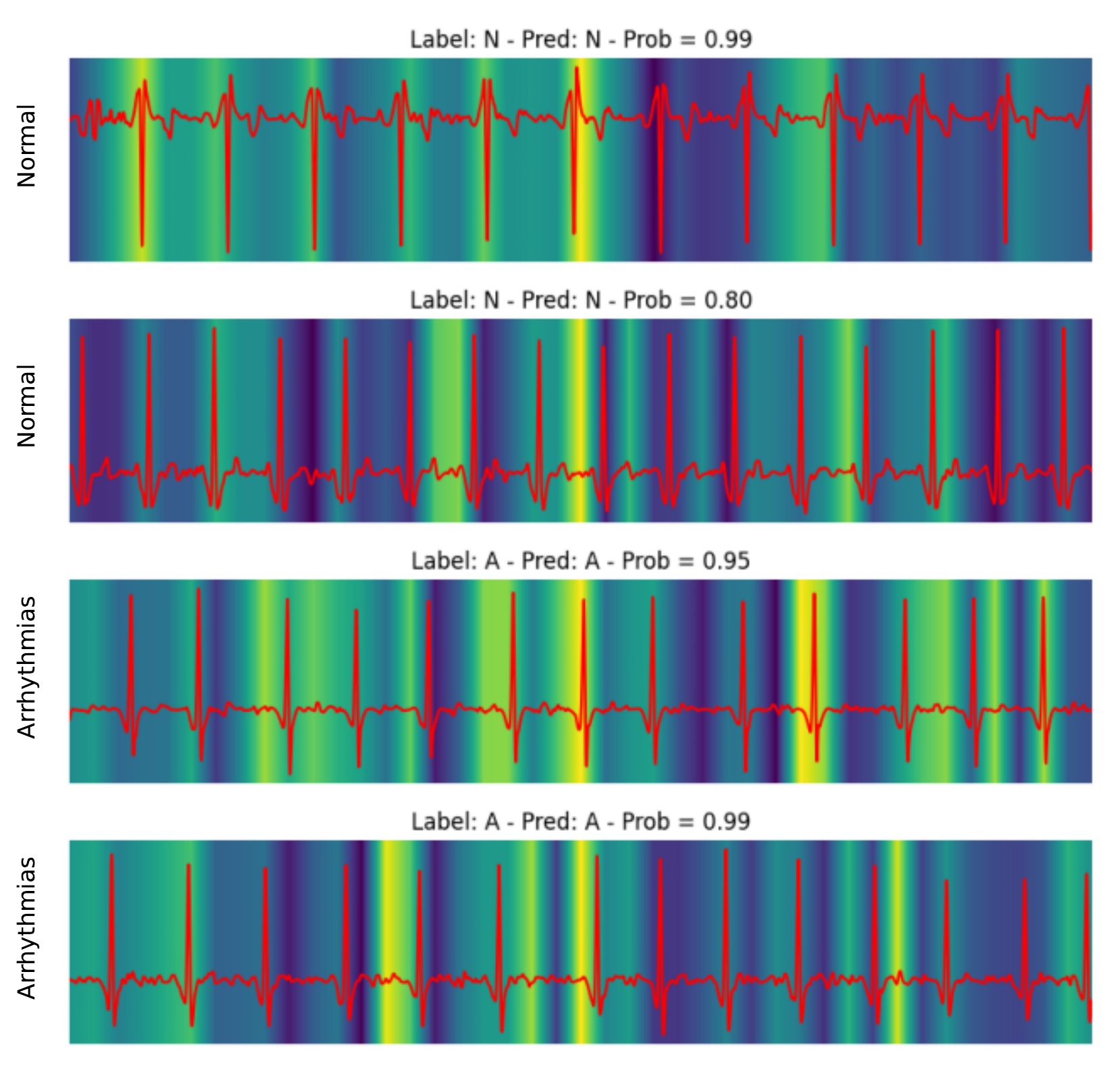}
    \caption{Explaining 1D ResNet decision by GradCam methods in case of normal regimes and arrhythmia}
    \label{fig:explain_resnet1d}
\end{figure}

Figure \ref{fig:explain_resnet1d} shows the focusing points of the 1D ResNet when predicting the AF signal. The yellow area is the segment that the model attracts. These heatmaps show that the classifier focused on the signal at the neighbor of the QRS complex. These regions are corresponding to the P-wave and T-wave of ECG recordings. In fact, the absence or abnormality of P-wave and T-wave is related to the fluctuation of heart rate and predicts arrhythmia disease \cite{hampton2013ecg}.

\subsubsection{Feature importance of XGBoost}

To explore how model XGBoost predicts classes, the feature importance score was calculated and summarized in Table~\ref{tab:fi_xgboost}. The results show that the features relating to the peak of signal, like \texttt{fft\_coefficient} and \texttt{ratio\_beyond\_r\_sigma}, are the highly important features. We can see that the XGBoost model infers the heart rate information indirectly via the peak-related features, after that the model could give the prediction of arrhythmia from heart rate.

\begin{table}[!ht]
   \caption{Feature importance score of each feature group in the model \texttt{XGBoost}} 
   \label{tab:fi_xgboost}
   \small
   \centering
   \begin{tabular}{p{0.25\linewidth} | p{0.1\linewidth}  p{0.15\linewidth} }
   \toprule\toprule
   \textbf{Group} & \textbf{\# features} & \textbf{Importance score}  \\ 
   \midrule

    \texttt{fft\_coefficient} & 22 & 0.2138 \\ 
    \texttt{ratio\_beyond\_r\_sigma} & 10 & 0.1989 \\
    \texttt{autocorrelation} & 8 & 0.0935 \\ 
    \texttt{energy\_ratio\_by\_chunks} & 5 & 0.0657 \\ 
    \texttt{index\_mass\_quantile} & 5 & 0.0651 \\ 
    \texttt{lempel\_ziv\_complexity} & 5 & 0.0483 \\ 
    \texttt{agg\_autocorrelation} & 3 & 0.0444 \\ 
    \texttt{range\_count} & 3 & 0.0411 \\ 
    \texttt{spkt\_welch\_density} & 3 & 0.0408 \\ 
    \texttt{change\_quantiles} & 3 & 0.0316 \\ 
    \texttt{quantile} & 2 & 0.0279 \\ 
    \texttt{number\_peaks} & 2 & 0.0241 \\ 
    \texttt{count\_below} & 1 & 0.0185 \\ 
    \texttt{cwt\_coefficients} & 2 & 0.0154 \\ 
    \texttt{number\_crossing\_m} & 1 & 0.0137 \\ 
       
   \bottomrule
  \end{tabular}
  
\end{table}

\subsection{Inference Time}
Table \ref{tab:infer_time} shows the comparison in the inference time among trained methods. Although that  XGBoost had a lightning prediction time, this model dominated the total inference time benchmark, which comes from the heavy processing steps. This problem leads to the fact that XGBoost still inferred 24 times longer than the second place. The Poincaré-based method requires an approximate two-fold longer inference time than the 1D CNN or 1D ResNet. This result complies with the mathematical characteristics of the 1D and 2D convolutional operators.

\begin{table}[!ht]
   \caption{The inference time of trained models.} 
   \label{tab:infer_time}
   \small
   \centering
   \begin{tabular}{p{0.12\linewidth} | p{0.13\linewidth}  p{0.13\linewidth}  p{0.11\linewidth}}
   \toprule\toprule
   \textbf{Model} & \textbf{Processing (ms)} & \textbf{Prediction (ms)} & \textbf{Total (ms)} \\ 
   \midrule
   
   \textbf{ResNet50} & 33.7 & 37.9 & 71.6  \\
   \textbf{DenseNet121} & 33.7 & 38.2 & 71.8 \\
   \textbf{1D CNN} & 13.5 & 27.5 & 41.0 \\
   \textbf{1D ResNet} & 13.5 & 18.8 & 32.2 \\
   \textbf{XGBoost} & 1717.6 & 0.2 & 1717.8 \\
   \bottomrule
  \end{tabular}
  
\end{table}

\subsection{Statement on computational resources and environmental impact}  

The experiment was performed on a workstation with 1 CPU Intel Core i7-9700F and 1 GPU NVIDIA RTX 3600. This work contributed totally 1.8 kg equivalent $CO_{2}$ emissions. The carbon emissions information was generated using the open-source library \textit{eco2AI}\footnote{Source code for \textit{eco2AI} is available at \url{https://github.com/sb-ai-lab/Eco2AI}} \cite{budennyy2023eco2ai}.

\section{Conclusion}
In the paper, we have presented novel various approaches to classify cardiac diseases from ECG recordings. The first approach took advantage of the Poincaré diagram and deep-learning-based image classifiers. ResNet50 and DenseNet121 architecture were chosen to process the graph. The experimental results figured out that these methods are decent for atrial fibrillation but not good at predicting other types of arrhythmia. In particular, the Poincaré-based methods have adequate performance in the CinC~2017 data but not good in the CinC~2020 data. However, RR or NN intervals, and therefore Poincaré diagrams, are much more accessible and can be obtained without the relatively complicated and expensive ECG procedure. Thus, it is still worth studying further in this approach. 
The XGBoost's performance is more impressive in the subset of long-term than the short-term data. This gradient-boosting model has a long inference time because of the expensive calculation in the preprocessing step. 
The one-dimensional convolutional model showed the best results in both studied datasets. Especially the 1D ResNet was superior to the first-ranking solution of each challenge. The residual connection showed its advantages in transferring information while keeping the model not too deep.

We have also investigated the efficiency metrics while training the models, including power consumption and equivalent CO2 emissions. Because of the high workload when processing 2D images, the 2D ResNet and DenseNet are at the top in power-consuming rankings. The XGBoost is energy efficient for the short term, but the power requirement is multiplied many times when training on long-term signals. Since the 1D convolution operator is optimized in the calculation, the unidimensional models like 1D CNN and 1D ResNet are the most energy efficient among the studied methods.

In the aspect of model interpretation, three models (DenseNet, 1D ResNet, and XGBoost) were analyzed to figure out how they discriminate the normal and AF data. The DenseNet detected AF using the heart rate variability, which was measured by the spreading of the data cloud and the presence of data in the upper-left and lower-right in the Poincaré diagram. On the other hand, the 1D ResNet assessed the AF pattern in raw ECG signal similar to a medical expert: this model focused on the area around the QRS complex, which is also the location of P and T waves.


\newpage
\bibliography{references}  

\begin{thebibliography}{28}
\providecommand{\natexlab}[1]{#1}
\providecommand{\url}[1]{\texttt{#1}}
\expandafter\ifx\csname urlstyle\endcsname\relax
  \providecommand{\doi}[1]{doi: #1}\else
  \providecommand{\doi}{doi: \begingroup \urlstyle{rm}\Url}\fi

\bibitem[Tsao et~al.(2022)Tsao, Aday, Almarzooq, Alonso, Beaton, Bittencourt,
  Boehme, Buxton, Carson, Commodore-Mensah, Elkind, Evenson, Eze-Nliam,
  Ferguson, Generoso, Ho, Kalani, Khan, Kissela, Knutson, Levine, Lewis, Liu,
  Loop, Ma, Mussolino, Navaneethan, Perak, Poudel, Rezk-Hanna, Roth, Schroeder,
  Shah, Thacker, VanWagner, Virani, Voecks, Wang, Yaffe, and Martin]{Tsao2022}
Connie~W. Tsao, Aaron~W. Aday, Zaid~I. Almarzooq, Alvaro Alonso, Andrea~Z.
  Beaton, Marcio~S. Bittencourt, Amelia~K. Boehme, Alfred~E. Buxton, April~P.
  Carson, Yvonne Commodore-Mensah, Mitchell~S.V. Elkind, Kelly~R. Evenson,
  Chete Eze-Nliam, Jane~F. Ferguson, Giuliano Generoso, Jennifer~E. Ho, Rizwan
  Kalani, Sadiya~S. Khan, Brett~M. Kissela, Kristen~L. Knutson, Deborah~A.
  Levine, Tené~T. Lewis, Junxiu Liu, Matthew~Shane Loop, Jun Ma, Michael~E.
  Mussolino, Sankar~D. Navaneethan, Amanda~Marma Perak, Remy Poudel, Mary
  Rezk-Hanna, Gregory~A. Roth, Emily~B. Schroeder, Svati~H. Shah, Evan~L.
  Thacker, Lisa~B. VanWagner, Salim~S. Virani, Jenifer~H. Voecks, Nae-Yuh Wang,
  Kristine Yaffe, and Seth~S. Martin.
\newblock Heart disease and stroke statistics—2022 update: A report from the
  american heart association.
\newblock \emph{Circulation}, 145, 2 2022.
\newblock ISSN 0009-7322.
\newblock \doi{10.1161/CIR.0000000000001052}.

\bibitem[Mensah et~al.(2019)Mensah, Roth, and Fuster]{mensah2019global}
George~A Mensah, Gregory~A Roth, and Valentin Fuster.
\newblock The global burden of cardiovascular diseases and risk factors: 2020
  and beyond, 2019.

\bibitem[Kligfield et~al.(2007)Kligfield, Gettes, Bailey, Childers, Deal,
  Hancock, van Herpen, Kors, Macfarlane, Mirvis, Pahlm, Rautaharju, and
  Wagner]{Kligfield2007}
Paul Kligfield, Leonard~S. Gettes, James~J. Bailey, Rory Childers, Barbara~J.
  Deal, E.~William Hancock, Gerard van Herpen, Jan~A. Kors, Peter Macfarlane,
  David~M. Mirvis, Olle Pahlm, Pentti Rautaharju, and Galen~S. Wagner.
\newblock Recommendations for the standardization and interpretation of the
  electrocardiogram.
\newblock \emph{Circulation}, 115:\penalty0 1306--1324, 3 2007.
\newblock ISSN 0009-7322.
\newblock \doi{10.1161/CIRCULATIONAHA.106.180200}.

\bibitem[Berkaya et~al.(2018)Berkaya, Uysal, Gunal, Ergin, Gunal, and
  Gulmezoglu]{Selcan2018}
Selcan~Kaplan Berkaya, Alper~Kursat Uysal, Efnan~Sora Gunal, Semih Ergin,
  Serkan Gunal, and M.~Bilginer Gulmezoglu.
\newblock A survey on ecg analysis.
\newblock \emph{Biomedical Signal Processing and Control}, 43:\penalty0
  216--235, 5 2018.
\newblock ISSN 17468094.
\newblock \doi{10.1016/j.bspc.2018.03.003}.

\bibitem[Clifford et~al.(2017)Clifford, Liu, Moody, wei Lehman, Silva, Li,
  Johnson, and Mark]{cinc2017}
Gari Clifford, Chengyu Liu, Benjamin Moody, Li~wei Lehman, Ikaro Silva, Qiao
  Li, Alistair Johnson, and Roger Mark.
\newblock Af classification from a short single lead ecg recording: the
  physionet computing in cardiology challenge 2017.
\newblock 9 2017.
\newblock \doi{10.22489/CinC.2017.065-469}.

\bibitem[Alday et~al.(2020)Alday, Gu, Shah, Robichaux, Wong, Liu, Liu, Rad,
  Elola, Seyedi, Li, Sharma, Clifford, and Reyna]{cinc2020}
Erick A~Perez Alday, Annie Gu, Amit~J Shah, Chad Robichaux, An-Kwok~Ian Wong,
  Chengyu Liu, Feifei Liu, Ali~Bahrami Rad, Andoni Elola, Salman Seyedi, Qiao
  Li, Ashish Sharma, Gari~D Clifford, and Matthew~A Reyna.
\newblock Classification of 12-lead ecgs: the physionet/computing in cardiology
  challenge 2020.
\newblock \emph{Physiological Measurement}, 41:\penalty0 124003, 12 2020.
\newblock ISSN 0967-3334.
\newblock \doi{10.1088/1361-6579/abc960}.

\bibitem[Hong et~al.(2020)Hong, Xu, Khare, Priambada, Maher, Aljiffry, Sun, and
  Tumanov]{Hong2020}
Shenda Hong, Yanbo Xu, Alind Khare, Satria Priambada, Kevin Maher, Alaa
  Aljiffry, Jimeng Sun, and Alexey Tumanov.
\newblock Holmes: Health online model ensemble serving for deep learning models
  in intensive care units.
\newblock pages 1614--1624. ACM, 8 2020.
\newblock ISBN 9781450379984.
\newblock \doi{10.1145/3394486.3403212}.

\bibitem[Ribeiro et~al.(2020)Ribeiro, Ribeiro, Paixão, Oliveira, Gomes,
  Canazart, Ferreira, Andersson, Macfarlane, Meira, Schön, and
  Ribeiro]{Ribeiro2020}
Antônio~H. Ribeiro, Manoel~Horta Ribeiro, Gabriela M.~M. Paixão, Derick~M.
  Oliveira, Paulo~R. Gomes, Jéssica~A. Canazart, Milton P.~S. Ferreira,
  Carl~R. Andersson, Peter~W. Macfarlane, Wagner Meira, Thomas~B. Schön, and
  Antonio Luiz~P. Ribeiro.
\newblock Automatic diagnosis of the 12-lead ecg using a deep neural network.
\newblock \emph{Nature Communications}, 11:\penalty0 1760, 12 2020.
\newblock ISSN 2041-1723.
\newblock \doi{10.1038/s41467-020-15432-4}.

\bibitem[Zhang et~al.(2021)Zhang, Yang, Yuan, and Zhang]{Zhang2021}
Dongdong Zhang, Samuel Yang, Xiaohui Yuan, and Ping Zhang.
\newblock Interpretable deep learning for automatic diagnosis of 12-lead
  electrocardiogram.
\newblock \emph{iScience}, 24:\penalty0 102373, 4 2021.
\newblock ISSN 25890042.
\newblock \doi{10.1016/j.isci.2021.102373}.

\bibitem[Attia et~al.(2019)Attia, Friedman, Noseworthy, Lopez-Jimenez, Ladewig,
  Satam, Pellikka, Munger, Asirvatham, Scott, Carter, and Kapa]{Attia2019}
Zachi~I. Attia, Paul~A. Friedman, Peter~A. Noseworthy, Francisco Lopez-Jimenez,
  Dorothy~J. Ladewig, Gaurav Satam, Patricia~A. Pellikka, Thomas~M. Munger,
  Samuel~J. Asirvatham, Christopher~G. Scott, Rickey~E. Carter, and Suraj Kapa.
\newblock Age and sex estimation using artificial intelligence from standard
  12-lead ecgs.
\newblock \emph{Circulation: Arrhythmia and Electrophysiology}, 12, 9 2019.
\newblock ISSN 1941-3149.
\newblock \doi{10.1161/CIRCEP.119.007284}.

\bibitem[Lima et~al.(2021)Lima, Ribeiro, Paixão, Ribeiro, Pinto-Filho, Gomes,
  Oliveira, Sabino, Duncan, Giatti, Barreto, Jr, Schön, and Ribeiro]{Lima2021}
Emilly~M. Lima, Antônio~H. Ribeiro, Gabriela M.~M. Paixão, Manoel~Horta
  Ribeiro, Marcelo~M. Pinto-Filho, Paulo~R. Gomes, Derick~M. Oliveira, Ester~C.
  Sabino, Bruce~B. Duncan, Luana Giatti, Sandhi~M. Barreto, Wagner~Meira Jr,
  Thomas~B. Schön, and Antonio Luiz~P. Ribeiro.
\newblock Deep neural network-estimated electrocardiographic age as a mortality
  predictor.
\newblock \emph{Nature Communications}, 12:\penalty0 5117, 8 2021.
\newblock ISSN 2041-1723.
\newblock \doi{10.1038/s41467-021-25351-7}.

\bibitem[Jun et~al.(2018)Jun, Nguyen, Kang, Kim, Kim, and Kim]{Jun2018}
Tae~Joon Jun, Hoang~Minh Nguyen, Daeyoun Kang, Dohyeun Kim, Daeyoung Kim, and
  Young-Hak Kim.
\newblock Ecg arrhythmia classification using a 2-d convolutional neural
  network.
\newblock 4 2018.

\bibitem[Bertsimas et~al.(2021)Bertsimas, Mingardi, and
  Stellato]{Bertsimas2021}
Dimitris Bertsimas, Luca Mingardi, and Bartolomeo Stellato.
\newblock Machine learning for real-time heart disease prediction.
\newblock \emph{IEEE Journal of Biomedical and Health Informatics},
  25:\penalty0 3627--3637, 9 2021.
\newblock ISSN 21682208.
\newblock \doi{10.1109/JBHI.2021.3066347}.

\bibitem[Corradi et~al.(2019)Corradi, Buil, Canniere, Groenendaal, and
  Vandervoort]{Corradi2019}
Federico Corradi, Jeroen Buil, Helene~De Canniere, Willemijn Groenendaal, and
  Pieter Vandervoort.
\newblock Real time electrocardiogram annotation with a long short term memory
  neural network.
\newblock pages 1--4. IEEE, 10 2019.
\newblock ISBN 978-1-5090-0617-5.
\newblock \doi{10.1109/BIOCAS.2019.8918723}.

\bibitem[Teplitzky et~al.(2020)Teplitzky, McRoberts, and
  Ghanbari]{Teplitzky2020}
Benjamin~A. Teplitzky, Michael McRoberts, and Hamid Ghanbari.
\newblock Deep learning for comprehensive ecg annotation.
\newblock \emph{Heart Rhythm}, 17\penalty0 (5, Part B):\penalty0 881--888,
  2020.
\newblock ISSN 1547-5271.
\newblock \doi{https://doi.org/10.1016/j.hrthm.2020.02.015}.
\newblock URL
  \url{https://www.sciencedirect.com/science/article/pii/S154752712030117X}.
\newblock Digital Health Special Issue.

\bibitem[Park et~al.(2009)Park, Lee, and Jeon]{Park2009}
Jinho Park, Sangwook Lee, and Moongu Jeon.
\newblock Atrial fibrillation detection by heart rate variability in poincare
  plot.
\newblock \emph{BioMedical Engineering OnLine}, 8:\penalty0 38, 12 2009.
\newblock ISSN 1475-925X.
\newblock \doi{10.1186/1475-925X-8-38}.

\bibitem[Lian et~al.(2011)Lian, Wang, and Muessig]{Lian2011}
Jie Lian, Lian Wang, and Dirk Muessig.
\newblock A simple method to detect atrial fibrillation using rr intervals.
\newblock \emph{The American Journal of Cardiology}, 107:\penalty0 1494--1497,
  5 2011.
\newblock ISSN 00029149.
\newblock \doi{10.1016/j.amjcard.2011.01.028}.
\newblock URL
  \url{https://linkinghub.elsevier.com/retrieve/pii/S0002914911003407}.

\bibitem[Zhang et~al.(2015)Zhang, Guo, Xi, Fan, Wang, Bi, and Wang]{Zhang2015}
Lijuan Zhang, Tianci Guo, Bin Xi, Yang Fan, Kun Wang, Jiacheng Bi, and Ying
  Wang.
\newblock Automatic recognition of cardiac arrhythmias based on the geometric
  patterns of poincaré plots.
\newblock \emph{Physiological measurement}, 36:\penalty0 283--301, 2 2015.
\newblock ISSN 1361-6579.
\newblock \doi{10.1088/0967-3334/36/2/283}.
\newblock URL \url{http://www.ncbi.nlm.nih.gov/pubmed/25582837}.

\bibitem[Bashar et~al.(2021)Bashar, Han, Zieneddin, Ding, Fitzgibbons, Walkey,
  McManus, Javidi, and Chon]{Bashar2021}
Syed~Khairul Bashar, Dong Han, Fearass Zieneddin, Eric Ding, Timothy~P.
  Fitzgibbons, Allan~J. Walkey, David~D. McManus, Bahram Javidi, and Ki~H.
  Chon.
\newblock Novel density poincaré plot based machine learning method to detect
  atrial fibrillation from premature atrial/ventricular contractions.
\newblock \emph{IEEE Transactions on Biomedical Engineering}, 68:\penalty0
  448--460, 2 2021.
\newblock ISSN 0018-9294.
\newblock \doi{10.1109/TBME.2020.3004310}.
\newblock URL \url{https://ieeexplore.ieee.org/document/9123697/}.

\bibitem[Carreiras et~al.(2015--)Carreiras, Alves, Louren\c{c}o, Canento,
  Silva, Fred, et~al.]{biosppypaper}
Carlos Carreiras, Ana~Priscila Alves, Andr\'{e} Louren\c{c}o, Filipe Canento,
  Hugo Silva, Ana Fred, et~al.
\newblock {BioSPPy}: Biosignal processing in {Python}, 2015--.
\newblock URL \url{https://github.com/PIA-Group/BioSPPy/}.
\newblock [Online; accessed <today>].

\bibitem[Zong et~al.(2003)Zong, Heldt, Moody, and Mark]{zong2003open}
W~Zong, T~Heldt, GB~Moody, and RG~Mark.
\newblock An open-source algorithm to detect onset of arterial blood pressure
  pulses.
\newblock In \emph{Computers in Cardiology, 2003}, pages 259--262. IEEE, 2003.

\bibitem[Hamilton(2002)]{hamilton2002open}
Pat Hamilton.
\newblock Open source ecg analysis.
\newblock In \emph{Computers in cardiology}, pages 101--104. IEEE, 2002.

\bibitem[He et~al.(2016)He, Zhang, Ren, and Sun]{he2016deep}
Kaiming He, Xiangyu Zhang, Shaoqing Ren, and Jian Sun.
\newblock Deep residual learning for image recognition.
\newblock In \emph{Proceedings of the IEEE conference on computer vision and
  pattern recognition}, pages 770--778, 2016.

\bibitem[Huang et~al.(2017)Huang, Liu, Van Der~Maaten, and
  Weinberger]{huang2017densely}
Gao Huang, Zhuang Liu, Laurens Van Der~Maaten, and Kilian~Q Weinberger.
\newblock Densely connected convolutional networks.
\newblock In \emph{Proceedings of the IEEE conference on computer vision and
  pattern recognition}, pages 4700--4708, 2017.

\bibitem[Selvaraju et~al.(2017)Selvaraju, Cogswell, Das, Vedantam, Parikh, and
  Batra]{Selvaraju2017}
Ramprasaath~R. Selvaraju, Michael Cogswell, Abhishek Das, Ramakrishna Vedantam,
  Devi Parikh, and Dhruv Batra.
\newblock Grad-cam: Visual explanations from deep networks via gradient-based
  localization.
\newblock In \emph{2017 IEEE International Conference on Computer Vision
  (ICCV)}, pages 618--626, 2017.
\newblock \doi{10.1109/ICCV.2017.74}.

\bibitem[Head et~al.(2021)Head, Kumar, Nahrstaedt, Louppe, and
  Shcherbatyi]{skopt2021}
Tim Head, Manoj Kumar, Holger Nahrstaedt, Gilles Louppe, and Iaroslav
  Shcherbatyi.
\newblock scikit-optimize/scikit-optimize, October 2021.
\newblock URL \url{https://doi.org/10.5281/zenodo.5565057}.

\bibitem[Hampton(2013)]{hampton2013ecg}
J.~Hampton.
\newblock \emph{The ECG Made Easy}.
\newblock Made Easy. Elsevier Health Sciences, 2013.
\newblock ISBN 9780702046414.
\newblock URL \url{https://books.google.com.vn/books?id=MXSSAAAAQBAJ}.

\bibitem[Budennyy et~al.(2023)Budennyy, Lazarev, Zakharenko, Korovin,
  Plosskaya, Dimitrov, Akhripkin, Pavlov, Oseledets, Barsola,
  et~al.]{budennyy2023eco2ai}
SA~Budennyy, VD~Lazarev, NN~Zakharenko, AN~Korovin, OA~Plosskaya, DV~Dimitrov,
  VS~Akhripkin, IV~Pavlov, IV~Oseledets, IS~Barsola, et~al.
\newblock Eco2ai: carbon emissions tracking of machine learning models as the
  first step towards sustainable ai.
\newblock In \emph{Doklady Mathematics}, pages 1--11. Springer, 2023.

\end{thebibliography}

\end{document}